# An Empirical Study on Team Formation in Online Games

Essa Alhazmi*, Sameera Horawalavithana*, Adriana Iamnitchi*, John Skvoretz†, Jeremy Blackburn‡
*Computer Science and Engineering, University of South Florida
{ealhazmi,sameera1,aii}@mail.usf.edu
†Department of Sociology, University of South Florida
skvoretz@usf.edu
‡Computer Science, University of Alabama at Birmingham
jblkburn@uab.edu

*Abstract*—Online games provide a rich recording of interactions that can contribute to our understanding of human behavior. One potential lesson is to understand what motivates people to choose their teammates and how their choices lead to performance. We examine several hypotheses about team formation using a large, longitudinal dataset from a team-based online gaming environment. Specifically, we test how positive familiarity, homophily, and competence determine team formation in Battlefield 4, a popular team-based game in which players choose one of two competing teams to play on. Our dataset covers over two months of in-game interactions between over 380,000 players. We show that familiarity is an important factor in team formation, while homophily is not. Competence affects team formation in more nuanced ways: players with similarly high competence team-up repeatedly, but large variations in competence discourage repeated interactions.

## I. INTRODUCTION

Teams are ubiquitous in modern societies and especially in commerce, business, and industry. The factors that impact team formation are of practical research interest. Online environments offer great potential for systematically exploring factors that impact team formation. Existing studies focus on team and play dynamics in online games [6], [1], [5]. Research in these on-line environments has its own challenges, not the least of which is the tracking of factors over time that could impact a player's choice of teammates.

The objective of this paper is to understand the interplay among three factors that impact team formation in online team-based games: 1) positive familiarity, 2) similarity, and 3) competence. We focus on how these factors shape an individual's choice of team in two-team competitive first-person shooter games.

Positive familiarity is the positive past performance with a teammate, which may translate into incentives to team up again in future encounters. Negative familiarity could also occur, and in this case negative past performance with a teammate translates in to a *disincentive* to team up again. Similarity is the sociological principle of homophily. The principle of homophily suggests that people seek out others of similar socio-demographic background for interaction. Finally, competence is the skill known to contribute to success in a team.

We investigate what factors affect team formation via observations of 60,410 "Battlefield 4" matches, played by 384,066 distinct players, on 63 servers, located in 7 different countries. The Battlefield series is one of the most popular first person shooter (FPS) multiplayer franchises in the world. Battlefield 4 is designed to support up to 64 players (32 on each team) by default, more than twice the number of players as other popular games like Call of Duty and Counter-Strike. By choosing to focus on this game environment, we are able to examine team formation not just at the team level, but also at "squad" level, which are small teams of up to 5 players who can coordinate more tightly.

Overall, we find that familiarity is an important factor in team formation, while similarity is not. Further, we discover that competence affects team formation in more nuanced ways: highly skilled players tend to team-up repeatedly, while large variations in competence discourages repeated interactions.

## II. RELATED WORK

The factors that make individuals choose particular teammates were studied in contexts as diverse as online gaming, software development, and education, based on surveys, observations, or digital records of interactions. Three factors commonly studied are positive familiarity, defined as the existence of previous positive experiences, competence (represented as expertise or reputation), and homophily.

Familiarity was shown to have an impact both on team formation and on team performance. In online gaming communities, Hudson et al. [12] showed that familiarity and team trust are positively correlated and they improve team performance. Waddell and Peng [20] showed that positive familiarity leads to repeated play, which leads to friendship. Mason and Clauset [15] found that players perform better when they play with friends, and individual performance is independent of team performance. In addition, it has been observed that players tend to be more ambitious in games when they have good cooperation with friends [19]. Good cooperation within the team leads to better performance [16] and is a stronger motivator than competition [17].

In geographically-distributed software teams [7], where the challenge is coordination among team members, performance was found to depend on two independent factors: competence (defined as familiarity with the task) and familiarity with the other members in the team. When team familiarity is weak, competence was shown to significantly improve performance.

These results confirm studies in offline environments that also show that familiarity is a factor in team formation and performance. For example, cooperation among participants in social care institutions in The Netherlands was found to grow with familiarity [3] and led to higher success rate.

Competence is an intuitive factor in team formation, yet it has been studied more as it relates to team success. In online gaming, Kim et al. [14] studied team congruency vs. individual proficiency in League of Legends (LoL), and discovered that individual proficiency has a bigger influence to team success. However, because LoL uses match-making algorithms for forming teams, competence could not be evaluated in the context of team formation.

Homophily was shown to be a relevant factor in team formation and success. In a recent study done on Massive Open On-line Courses (MOOC) [6], Eftekhar et al. found that age, education level, distance and time zones are factors for students to form successful teams. Moreover, multidisciplinary teams with more diverse skill sets were more successful than the rest. Kamel et al. [13] viewed homophily as a distance function between different user skills and proposed it as a strategy for forming software development teams.

Few studies considered all three categories. By observing a group of students forming 33 small teams over four years, Hinds et al. [10] show how competence, familiarity and homophily affect team formation. Positive familiarity is shown to be correlated with competence: the better reputation among their peers a student has, the more often will be included in the same team. The study also shows how homophily plays a role in team formation: students show strong preference for working with colleagues of the same race, but less so for same gender. At the same time, homophily is the weakest factor.

Ruef et al. [18] studied a sample of 816 organizational founding teams and used structural event analysis to predict the number of entrepreneurial teams. The study shows that homophily and the existence of a previous relationship (friendship or family ties) are important factors to predict team composition, while spatial and geographical proximity are not.

Familiarity, competence and homophily are thus the main categories that were studied as determinants for team formation and performance. However, each context represents each of these categories by a different set of environment-specific variables. To our knowledge, this study is the first large-scale quantitative study on the formation of teams in online gaming community based on the combination of familiarity, homophily and competence.

## III. DATASET

We analyzed rich digital records of team dynamics in online games. In contrast with previous studies that looked at optional teaming up for task-specific objectives [11], we focused on team-based games, described below, where players must choose one team (out of typically two teams) in order to play. We collected data by observing GameMe, a service that provides statistical player information on a collection of games. In this work we narrowed our analysis on one highly popular game, Battlefield 4, because of its popularity and its use of squads as mini-teams.

### A. Team-Based Online Games

There are two high-level methods of team formation in online games: 1) matchmaking and 2) player choice. Matchmaking systems construct teams in such a way as to provide a 50% chance of each team to win. Matchmaking systems are typically used by popular eSports games like the League of Legends and Dota 2, and the competitive mode of Counter-Strike. Player-choice systems provide players with many more options ranging from the particular server (population) they play on, what team they want to join, and sometimes even down to a smaller "squad" level unit. We focus in this work on games with player-choice systems since matchmaking systems are not suitable for understanding how humans *choose* to be on a team. We describe how they work in further detail next.

**Player Connection.** Player choice systems are usually built around a "server browser," which lets players discover available servers by providing information like the map that is being played, the players actively playing on the server, and perhaps the scores of the players. Unlike matchmaking systems, players are free to connect to which ever server they want, and thus the server browser presents the first point at which humans are making a choice about team formation.

**Maps & Rounds.** Most multiplayer games take place on one or more *maps*. A map is a simulated environment in which the game takes place. For many games, and particularly the one that we study in this paper, there are multiple maps available. Maps vary not only in their layout and visual design, but also in the goals that players must meet. For example, one map might be a simple "death match" where the winning team is the one that kills more opposing players, while another map might be "point control" where the winning team is able to capture and control certain areas on the map.

Maps are usually rotated after some win condition or a time limit is reached. Because certain map types are asymmetrical (i.e., one team attacks and the other defends), many maps are broken down further into rounds. We use match and map synonymously in this paper, i.e., a match takes place on a single map, and has a beginning, and an end.

### B. GameME

GameME is a third-party service that monitors real-time playing activities for about 20 team-based on-line games on many gaming servers around the world. It provides real-time statistics on players' scores for the games monitored. GameME provides APIs to access two different views of the player population: a global view, that ranks players globally over all games monitored, and a local, server-specific view, that ranks players on the local game server.

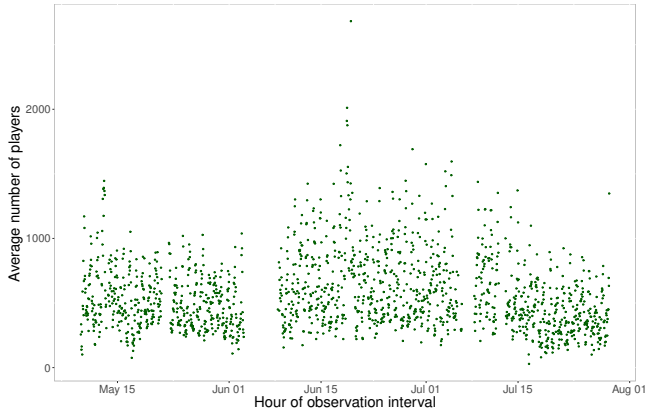

Fig. 1: The number of distinct Battlefield 4 players observed over the observation period (81 days).

The local view, accessible by the so-called "client API", provides detailed information about the current match played on each game server. This information includes attributes such as the name of the map played, the start time of the match, the players in each team and their squad membership, skill change in match, and current rank on the server. The client API also provides access to on-server player profiles that includes historical information (such as achievements, rewards, statistics on team membership), weapons owned and their usage, clan membership, reported location (country), etc.

*C. Data Collection*

We collected data from GameME between May 9, 2016 and July 29, 2016 (81 days). For this, we built a web crawler by using the GameMe Client API for all 20 games monitored. In order to distribute the load, we used 20 crawlers on five machines, with one crawler per game. For load balancing, we assigned varying loads to machines according to the popularity of the games. For the very popular games, the crawler took up to 50 minutes to record information about all active game servers, while for the less popular games the crawling took under a minute. We varied the periodicity of the crawling accordingly. We stored the information in a MongoDB database.

We cleaned the dataset by removing records for BOT players that are automatically generated to simulate activity on game servers. In addition, we removed the players who were not actively playing at the time of observation but appeared as spectators or unassigned in a team. As a result, we removed about 2.9 million records of BOT players and about 3.4 million records of spectator players (out of more than 32.5 million records).We ended up with 26.61 million records about 2.7 million distinct players over 81 days from 20 different games supported on 1,426 servers in 32 countries. For each player, we recorded more than 70 attributes that describe real time playing status and profile information.

*D. Battlefield 4*

In this study we focus on only Battlefield 4. Battlefield 4 is a very popular multiplayer first-person shooter game (FPS), played on a simulation of modern battlefield, and has features like a wide variety of weapons, maps, vehicles, and destructible terrain. The Battlefield series is known for large maps with a great number of players: Battlefield is designed around having over twice the number of players per match than typical FPSs.

This focus on large-scale games has resulted in some features that make Battlefield an excellent source for studying team formation. By its very nature, teams in Battlefield are generally too large (up to 32 players on each team by default) to operate as a single unit. To alleviate this, the developers included the *squad* concept.

A squad is a smaller unit of players that are on the same team. Squads can be up to 5 players: 1 leader and 4 other members. Battlefield 4 employs "classes" where certain abilities and weapons are restricted to each class. Thus, it is often wise to build a squad from a diverse set of classes to ensure that any challenges can be met. For example, it is a good idea for a squad to include not only assault classes, but also a medic class to heal and revive teammates as well as an engineer class to repair any vehicles the squad might come across.

Battlefield squads have very concrete advantages. First, the squad leader is able to issue orders using an in-game interface, such as attack or defend a given objective. When squad members successfully execute these orders, bonus points are received. Next, there are certain squad "specializations" that act as bonuses for the entire squad and are unlocked as the squad performs well throughout a match. Squads are also given their own chat channel (text and voice) which allows for increased communication. Finally, squad members can "respawn" (come back to life after dying) on top of any of their squad members instead of at a predefined spawn point. This allows fast redeployment to the action.

Our data shows that Battlefield 4 has 384,066 distinct players played in 60,410 matches on 63 servers located in 7 countries. The maximum team size in our dataset is 130 players. Figure 1 shows the number of players observed in Battlefield 4 over time. The gaps in crawling (about 7 days total) are due to temporary failures in data collection.

*E. Data Attributes*

Players have multiple profiles, one associated with each Battlefield game server they played on. Players are identifiable by their globally unique Battlefield ID, yet have a local (server-based) player ID on each server they access. Our dataset for Battlefield 4 shows that about 50% of players have only one profile, 40% of players have between two and five profiles. The maximum number of profiles per player in our dataset is 24. Figure 2 shows the distribution on the number of profiles per player.

Typical attributes used for identifying homophily such as gender, education, race, age, are not reported in the player pro-files. Moreover, we know that gender is often misrepresented in profiles of online gamers, due to toxicity and harassment [8], [4]. For homophily, we identified two potentially relevant attributes: country of origin and clan. The country of origin is declared by the player. Less than 1% of the players do not report a country of origin in any of their server profiles while about 9% of the players declare multiple countries (with a maximum of 5) in their different profiles However, 90% of players declare one country of origin, so we concluded that this is a meaningful data attribute for our analysis.

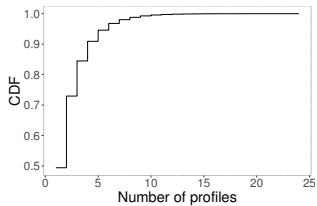

Fig. 2: Cumulative distribution function of the number of servers (i.e. number of profiles per player) where players appear during the observation period.

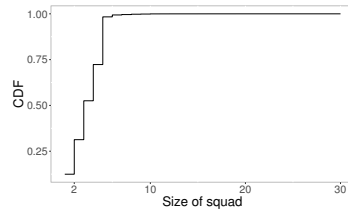

Fig. 3: Cumulative distribution function of squad size.

Clan is a free text tag that the player can choose to associate with other players. Unlike guilds [5] in Massively Multiplayer Online Games (such as War of Warcraft), clan membership has no influence on playing mechanics, but has only identification purposes. In Battlefield, as it turns out from our observations, clan membership does not have much importance on player identity, either: few players declare their belonging to a clan. Over 80% of the players do not declare a clan in their profile, less than 9% belong to one clan, and the remaining declare membership to more than one clan. We thus ignored this attribute from our analysis.

Many attributes in player profiles measure aspects of competence over time, such as number of kills, number of kills per minute, number of deaths, etc. However, all these measures can be uniquely represented by the server-based rank of a player, which is computed by sorting the players in decreasing order based on the number of points accumulated. Players start with a fix amount of (1000) points and acquire more when they "kill" opponents, and lose points when they are "killed". The number of points lost or gained depend both on the opponent's relative competence level (such that, for example, being killed by a newbie is more costly than being killed by an advance player) and on the weapon used (such that, for example, killing using a rifle gains fewer points than killing by knife). Players can thus have different ranks on the different servers they play. In our analysis, we used the rank of the player on the particular server on which we observed an interaction, as reported in that server's player profile at the time of the observation.

Two crucial attributes are useful for studying team membership. The first is the membership to one of two teams, typically called "US Army" and "Chinese Army" or "Russian Army". The second is membership to squads, with predefined names such as Alpha, Bravo, Charlie, Delta.

### F. Data Characterization

We ended up with 384,066 distinct players who played in 60,410 matches on 63 servers located in 7 countries. These players collectively report to come from 172 countries across the globe, spanning all continents. Over the observation period we counted 1,577,946 squads with at least 2 members each (we removed about 8% of the squads observed who were formed of only one player). As visible in Figure 3, the vast majority of the squads are of 5 or fewer players, but we notice (a very small percentage of) squads of unexpected size (up to 30 players). We suspect these are due to unusual server configurations that allow such large squads.

We observed that the majority player population is from the US (Table I). Surprisingly, the players from the country where the server is located do not always make the majority population on that server. For example, Russian players are the majority on the German servers, while German players form the majority on servers located in UK. However, geographical proximity, as expected, plays an important role in the players' choice of servers (due to latency considerations): servers attract the majority population from the same continent.

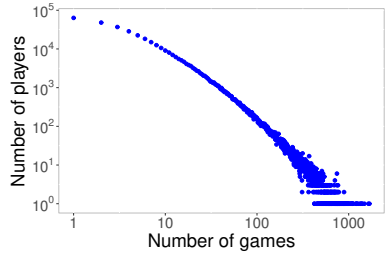

Fig. 4: Distribution of the number of matches per player.

One of the pre-requisites for any dataset used to understand team formation is that it provides plentiful samples of the same player making a decision on which team to join. To that end, Figure 4 plots the distribution of the number of games we observed for each player. The distribution is quite heavy tailed: while the mean and median number of games played are 5 and 17, about 2.7% of players played over 100 games while the most active player participating in 1682 games. Overall, the figure provides evidence that we have plentiful data to understand player preferences with team formation.

Figure 5 plots the distribution of the average number of active players observed per hour of the day for each day of the week (GMT timezone). We see expected patterns both in terms of which days and times the players are the most active: players are more active on non-workdays, with peaks occurring in the evenings (biased towards the US time zones) and troughs during typical working/school hours. Although these results are to be expected, they serve to validate our collection methodology.

TABLE I: Country of origin of the majority of players vs. server location.

| Server location (# players) (% population) | First Majority | Second Majority | Third Majority |
|---|---|---|---|
| United States (189,088) (61.66%) | United States (67.25%) | Canada (11.32%) | Brazil (9.15%) |
| Germany (52,990) (17.28%) | Russian Federation (24.5%) | Poland (21.2%) | Sweden (7%) |
| United Kingdom (34,464) (11.24%) | Germany (53%) | United Kingdom (22.6%) | France (17.5%) |
| Australia (25,584) (8.34%) | Australia (83.22%) | New Zealand (16.5%) | New Caledonia (0.25%) |
| Brazil (2,765) (0.9%) | Argentina (59%) | Chile (41%) | – |
| Netherlands (1,317) (0.43%) | Switzerland (100%) | – | – |
| Poland (462) (0.15%) | Estonia (85%) | Luxembourg (11.48%) | Montenegro (3.46%) |

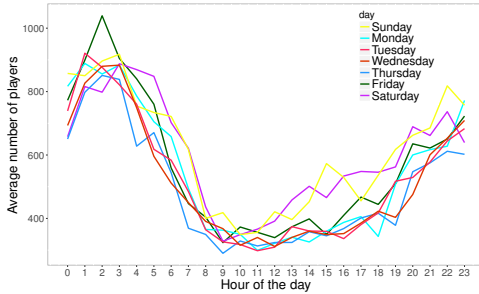

Fig. 5: Average number of players per-hour, per-day of the week (from Sunday) for Battlefield 4 (GMT time).

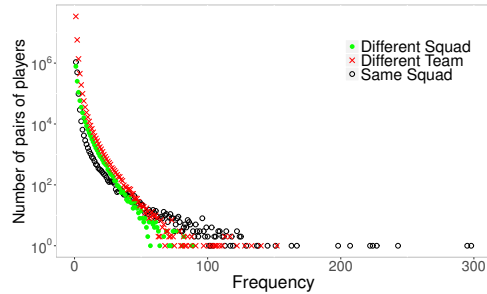

Fig. 6: Frequency distribution of pairs of players over the games played on the same, different squad, and different team.

## IV. Understanding Team Formation

In orders to understand team formation, we also focus on analyzing the formation of squads, for reasons already mentioned. Specifically, given the large difference in size between teams and squads in Battlefield 4, squads are more likely to represent meaningful in-team collaboration, affinity and choice, that can be translated into understandings about team formation in contexts other than online games.

We structure our analysis around the three factors for team formation presented before: familiarity, homophily and competence.

### A. Familiarity

Familiarity affects the choice to join a particular squad under the intuition that positive past experiences with another player will result in teaming up in the future. However, the question remains whether familiarity has any meaningful effect with respect to team formation in online gaming environments. Unlike the real world, players of online video games typically have *millions* of other players to play with.

Figure 6 plots the distribution of pairs of players with respect to the number of times they were on the same squad, different squads, and different teams. In general, we see a preference towards familiarity: players become increasingly likely to be on the same squad as the number of games played together increases. Conversely, it becomes increasingly unlikely that "familiar" players will choose to be on different squads and teams.

Recall that there are concrete benefits to being in the same squad, and this is likely one of the reasons why players who become more familiar with each other tend to play on the same squad repeatedly as shown in Figure 7. Once they have played enough together on the same team and start to understand each others' play style, it only makes sense to join up together. The players can earn more points and get squad perks, as well as having their own private communication channel. (Note that a squad is by definition part of the same team.)

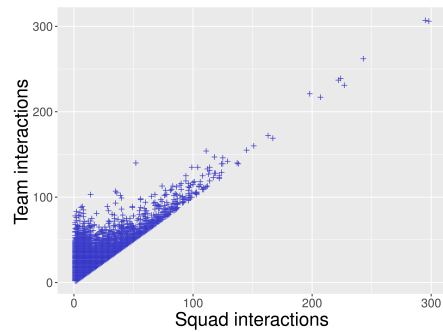

Fig. 7: Number of co-team vs. co-squad for pairs of players.

### B. Homophily

Homophily, or the tendency for similar people to form relationships, was shown to have a role in the formation of student teams for class projects [10]. In the gaming context, homophily with respect to the country of origin is intuitive due to shared language and solidarity against a shared political enemy. Moreover, cultural characteristics were shown to play a role in the choice of people to play [9]. We thus look at shared country of origin for members of the same squad or the same team. First we analyze the distribution of pairs of players who played on the same team.

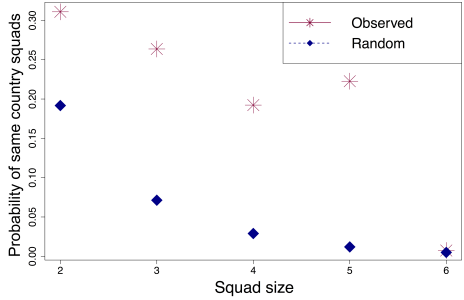

Fig. 8: Probability distribution of squad formation over the players in the same country and randomly assigned country.

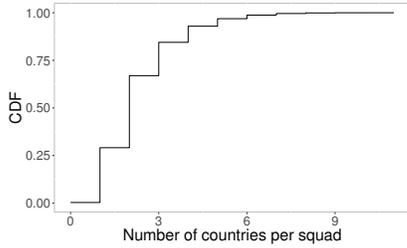

Fig. 9: CDF of squad diversity over the country of origin.

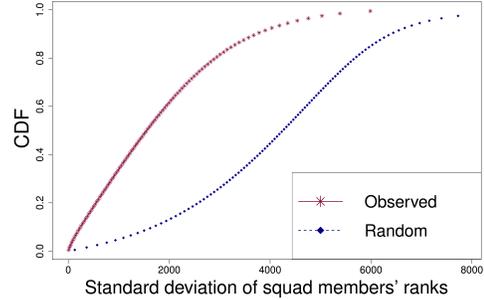

Fig. 10: CDF of standard deviation of squad members' ranks compared with the randomly generated squads. Note that players have distinct ranks and the squads in the random assignment follow the same size distribution as the real dataset.

Table I shows server locations and the top three countries of players that play on them. Generally speaking, we can see that servers are populated by players from "near by" countries. Figure 8 compares the probability distribution of squad formation for players from the same country with a random distribution, which assigns players to squads based on the population of player origins. From the figure, we see that the smaller squads are more likely than larger ones to be composed of players all from the same country. If homophily with respect to country of origin were a major factor in team formation, we would expect the number of larger squads with only a single country to be heavily biased towards the majority country of the population. However, we see that a two-thirds of squads is formed by at least two countries (Figure 9).

### C. Competence

The theory behind competence and team formation is that players will naturally gravitate towards players that are "good" at the game. This makes intuitive sense: everyone wants to be on a team with a winner. However, competency with respect to team formation is a double-edged sword: although everyone might want to team up with the best player, that best player wants to team up with good players, not just any random person. Moreover, the notoriously toxic and unforgiving environment of online gaming makes it potentially unpleasant to play with much better players: the less competent player risks nasty comments.

In this section we study how competency affects team formation. We considered the competence of a player as described by his/her rank on the server on which the playing is observed. We are particularly interested in divining the interplay between wanting to play with the best and wanting to play with players with similar competence.

Figure 10 plots the CDF of the standard deviation of squad members' ranks compared to squads that are randomly populated. For the random squad generation, we followed the distribution of squad sizes from the real dataset and randomly assigned ranks to the squad members. Since ranks are unique on a server, once a rank was assigned, it cannot be assigned again to another squad. From the figure, we observe that the standard deviation of ranks from real squads are much lower than those from the randomly generated squads. The median standard deviation for real squads is around 1,800 whereas it is over 4,000 for the randomly generated squads. We further note that 90% of real squads have a standard deviation *less than* the median of randomly generated squads.

What this means is that players are definitively *not* joining squads at random, but rather joining them in a manner that suggests competency as an underlying metric. I.e., players are naturally forming squads with players that are around their same competency level.

### V. SQUAD MEMBERSHIP PREDICTION

In this section, we build predictive models based on our analysis in the previous section. Out of 63 Battlefield-4 servers, we focus on the four most popular in terms of distinct players and number of games played. We extract ten features that belong to homophily, familiarity, and competence categories.

Our goal is to build a prediction model to determine whether a pair of players in a game will join the same squad. We select a subset of data for each feature based on their characteristic distribution, considering category representation and size of the resulting dataset. After that, we evaluate different models and choose the best classifier for the most popular server. Finally, we evaluate the accuracy of the model trained on the most popular server, by testing on the dataset from the other three servers.

*A. Features and Categories*

Since our focus is to predict whether a pair of players will join the same squad, we build different models based on the three categories that influence team formation. Same clan, same country (both binary), and rank distance (RD) (numeric) are three features for pairs that reflect the homophily category. For competence, we extract four numeric features that formulate pairs performance in a game: average rank (AR), average skill change rate (ASCR), average kills per deaths (AKPD), and average head shots (AHS).

In the preliminary process of building our models, we split the pairs of players on each server into two sets: the training set includes the first 60 days of our dataset, while the testing set contains the remaining 21 days. We consider same side team frequency (SST) (i.e., the players were on the same team during the game) and different side team frequency (DST) for the first 60 days as features belonging to the category of familiarity. The combination of SST and DST defines the frequency of appearing in the same game.

Because the player activity and attributes are highly variable, we select subpopulations with similar characteristics such that we can better map them into categories. For example, we select the players with average rank (AR) $< 50$, thus with high competence level. Our choice of thresholds (e.g., 50 for AR and RD in Table II) is informed by the distribution of the features and the objective of selecting at least 10,000 pairs of players above that threshold.

Our dataset is unbalanced, with the majority of pairs from different squads and only approximately 10% from the same squad. We use the synthetic minority over-sampling technique (SMOTE) [2] algorithm to create a balanced dataset from the unbalanced one in the training set. The Area Under the Curve (AUC) and Accuracy measures provide a summary of the quality of the classifier.

*B. Model Evaluation and Results*

The accuracy and AUC of various classifiers for our models can be seen in Table II. The Random Forest classifier performs better than other classifiers in both metrics on all subsets of features. Moreover, we find that familiarity features are the most influential features. The accuracy and AUC of the set of data controlled for SST exceed 80%. For DST, accuracy exceeds 80% and AUC exceeds 70%.

Next, we evaluate how different sampling strategies affect the results of our models based on different thresholds. We have two types of models: the local models are trained and tested from the data drawn from each server, and the generalized model is trained only from the most popular server, but tested on each server's local population.

As shown in Figure 11, the upper left plot demonstrates the quality of each local model as we vary the threshold of the number of games played together. The local models perform quite well if we choose our threshold to be between 50 and 80 games played together. For the number of games played together, we used SST+DST as the familiarity feature. The remaining plots show how the generalized model compares to the local models. By comparing two predictive models; local and global which are controlled by the frequency of games, we found that the generalized predictive models perform similar to local model in server 2, but outperform it in server 3 and 4. Overall, our findings indicate that we can discover a global model from the most popular server.

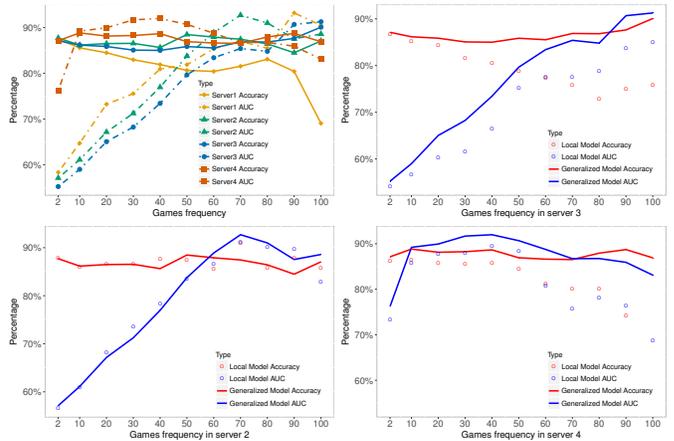

Fig. 11: Local predictive models compared to generalized predictive models in squad formation using random forest classifier on different frequencies of games, which includes same side and different side teams, for all pairs of players in the four most popular servers.

## VI. SUMMARY AND DISCUSSIONS

Online environments offer rich potential for studying how individuals choose teammates for a goal-oriented activity. Our results provide evidence consistent with the factors identified by Hinds et al. [10] as important in team formation: familiarity, homophily, and competence. We find that the more often two players play together, the more likely it is that they are playing on the same (large) team and even on the same (smaller) squad within that team. We presume that this effect occurs in part because pairs that had success in previous interactions continue to select one another as partners while those that failed in previous interactions differentially drop out of future association. A more fine grained dynamic examination of partner choice that would condition on previous game outcome is needed to substantiate this interpretation.

Homophily as a factor clearly shows up with respect to country. To be fair, this fact could be due simply to a time/distance difference between players in different countries serving as a disincentive to play together at the same time on a server in one or another's country, especially considering that distance can heavily affect latency. Nevertheless, the extent to which interactions between those in the same country exceed chance expectations is quite large, and not just for dyadic teammate relations but also when we examine the composition of squads of different sizes. Homophily also occurs with respect to skill level. Our results make it clear that teammates and squad mates are far more likely to be close to each other in skill than could be expected by chance.

Please cite the ASONAM17 versionTABLE II: A comparison of five classifiers for squad prediction over homophily, competence, and familiarity as categories of features. a) Pairs of players in the same clan represent 0.075% of all pairs, a very small proportion of our dataset, however this set includes 66% of the pairs in the same squad and 76% of the pairs in same team. b) Pairs of players in the same country represent about 59% of all pairs. However, this set includes only 5.2% of the pairs in the same squad and 49% in the same teams.

| Classifiers | | Random Forest | | AdBoost | | Decision Tree | | Naive Bayes | | Logistic Reg. | |
|---|---|---|---|---|---|---|---|---|---|---|---|
| Categories | Features | Accuracy | AUC | Accuracy | AUC | Accuracy | AUC | Accuracy | AUC | Accuracy | AUC |
| **Homophily** | *Same clan* | 77.08 | 68.32 | 72.65 | 70.44 | 66.00 | 54.43 | 54.37 | 55.04 | 49.72 | 71.18 |
| | *Same country* | 89.49 | 52.22 | 89.11 | 52.93 | 80.37 | 50.83 | 88.43 | 52.79 | 87.76 | 53.44 |
| | *RD<50* | 85.92 | 60.49 | 85.33 | 59.50 | 74.92 | 53.61 | 82.44 | 57.37 | 82.11 | 58.25 |
| **Competence** | *AR<50* | 78.26 | 61.69 | 76.5 | 64.81 | 68.42 | 55.77 | 74.92 | 59.81 | 76.05 | 60.78 |
| | *ASCR>5* | 89.79 | 52.69 | 89.62 | 56.49 | 81.59 | 51.04 | 89.59 | 56.44 | 88.61 | 53.65 |
| | *AKPD>7* | 89.72 | 53.67 | 89.4 | 53.26 | 81.87 | 50.74 | 71.77 | 54.43 | 85.14 | 53.50 |
| | *AHS>7* | 88.28 | 54.53 | 88.44 | 56.70 | 81.00 | 52.90 | 86.38 | 51.12 | 87.57 | 54.11 |
| **Familiarity** | *SST>20* | 80.59 | **81.57** | 78.60 | 81.52 | 69.77 | 68.15 | 70.19 | 74.89 | 79.08 | 77.55 |
| | *DST>20* | **81.77** | 72.61 | 75.65 | 69.00 | 70.15 | 57.85 | 72.74 | 66.22 | 64.15 | 68.20 |

Finally, it is also clear that competence is a driver for interaction, but the connection is complex. First, we found that the most frequent pair interactions occur between players on the same squad with relatively high rank (i.e., very competent players). Likewise, pairs that are on average *less* competent have less frequent interaction. Second, we saw a similar story with respect to the absolute difference in rank between the pairs. I.e., players that were of approximately the same rank were increasingly likely to have more interactions together.

ACKNOWLEDGMENT

This research was supported by the U.S. National Science Foundation under Grants No. IIS 1546453. The first author was supported by a scholarship from Jazan University.REFERENCES

[1] G. A. Benefield, C. Shen, and A. Leavitt. Virtual team networks: How group social capital affects team success in a massively multiplayer online game. In *Proceedings of the 19th ACM Conference on Computer-Supported Cooperative Work & Social Computing*, CSCW '16, pages 679–690, New York, NY, USA, 2016. ACM.

[2] N. V. Chawla, K. W. Bowyer, L. O. Hall, and W. P. Kegelmeyer. Smote: synthetic minority over-sampling technique. *Journal of artificial intelligence research*, 16:321–357, 2002.

[3] A. C. Costa, R. A. Roe, and T. Taillieu. Trust within teams: The relation with performance effectiveness. *European journal of work and organizational psychology*, 10(3):225–244, 2001.

[4] A. C. Cote. i can defend myself. *Games and Culture*, 0(0):1555412015587603, 0.

[5] N. Ducheneaut, N. Yee, E. Nickell, and R. J. Moore. "alone together?": Exploring the social dynamics of massively multiplayer online games. In *Proceedings of the SIGCHI Conference on Human Factors in Computing Systems*, CHI '06, pages 407–416, New York, NY, USA, 2006. ACM.

[6] M. Eftekhar, F. Ronaghi, and A. Saberi. Team formation dynamics: A study using online learning data. In *Proceedings of the 2015 ACM on Conference on Online Social Networks*, COSN '15, pages 257–267, New York, NY, USA, 2015. ACM.

[7] J. A. Espinosa, S. A. Slaughter, R. E. Kraut, and J. D. Herbsleb. Familiarity, complexity, and team performance in geographically distributed software development. *Organization Science*, 18(4):613–630, July 2007.

[8] J. Fox and W. Y. Tang. Womens experiences with general and sexual harassment in online video games: Rumination, organizational responsiveness, withdrawal, and coping strategies. *New Media & Society*, 0(0):1461444816635778, 2016.

[9] C. E. Goodfellow. Online gaming in post-soviet russia: Practices, contexts and discourses. 2015.

[10] P. J. Hinds, K. M. Carley, D. Krackhardt, and D. Wholey. Choosing work group members: Balancing similarity, competence, and familiarity. *Organizational behavior and human decision processes*, 81(2):226–251, 2000.

[11] Y. Huang, W. Ye, N. Bennett, and N. Contractor. Functional or social?: Exploring teams in online games. In *Proceedings of the 2013 Conference on Computer Supported Cooperative Work*, CSCW '13, pages 399–408, New York, NY, USA, 2013. ACM.

[12] M. Hudson, P. Cairns, and A. I. Nordin. *Familiarity in Team-Based Online Games: The Interplay Between Player Familiarity and the Concepts of Social Presence, Team Trust, and Performance*, pages 140–151. Springer International Publishing, Cham, 2015.

[13] K. Kamel, Z. Al Aghbari, and I. Kamel. Realistic team formation using navigation and homophily. In *2014 International Conference on Big Data and Smart Computing (BIGCOMP)*, pages 197–203. IEEE, 2014.

[14] J. Kim, B. C. Keegan, S. Park, and A. Oh. The proficiency-congruency dilemma: Virtual team design and performance in multiplayer online games. In *Proceedings of the 2016 CHI Conference on Human Factors in Computing Systems*, CHI '16, pages 4351–4365, New York, NY, USA, 2016. ACM.

[15] W. A. Mason and A. Clauset. Friends ftw! friendship and competition in halo: Reach. *CoRR*, abs/1203.2268, 2012.

[16] W. Peng and G. Hsieh. The influence of competition, cooperation, and player relationship in a motor performance centered computer game. *Computers in Human Behavior*, 28(6):2100–2106, 2012.

[17] N. Pobiedina, J. Neidhardt, M. d. C. Calatrava Moreno, and H. Werthner. Ranking factors of team success. In *Proceedings of the 22Nd International Conference on World Wide Web*, WWW '13 Companion, pages 1185–1194, New York, NY, USA, 2013. ACM.

[18] M. Ruef, H. E. Aldrich, and N. M. Carter. The structure of founding teams: Homophily, strong ties, and isolation among u.s. entrepreneurs. *American Sociological Review*, 68(2):195–222, 2003.

[19] M. Spante, I. Heldal, A.-S. Axelsson, and R. Schroeder. Strangers and friends in networked immersive environments: Virtual spaces for future living. 2003.

[20] J. C. Waddell and W. Peng. Does it matter with whom you slay? the effects of competition, cooperation and relationship type among video game players. *Computers in Human Behavior*, 38:331–338, 2014.
8

TABLE II: A comparison of five classifiers for squad prediction over homophily, competence, and familiarity as categories of features. a) Pairs of players in the same clan represent 0.075% of all pairs, a very small proportion of our dataset, however this set includes 66% of the pairs in the same squad and 76% of the pairs in same team. b) Pairs of players in the same country represent about 59% of all pairs. However, this set includes only 5.2% of the pairs in the same squad and 49% in the same teams.

| Classifiers | | Random Forest | | AdBoost | | Decision Tree | | Naive Bayes | | Logistic Reg. | |
|---|---|---|---|---|---|---|---|---|---|---|---|
| Categories | Features | Accuracy | AUC | Accuracy | AUC | Accuracy | AUC | Accuracy | AUC | Accuracy | AUC |
| **Homophily** | *Same clan* | 77.08 | 68.32 | 72.65 | 70.44 | 66.00 | 54.43 | 54.37 | 55.04 | 49.72 | 71.18 |
| | *Same country* | 89.49 | 52.22 | 89.11 | 52.93 | 80.37 | 50.83 | 88.43 | 52.79 | 87.76 | 53.44 |
| | *RD<50* | 85.92 | 60.49 | 85.33 | 59.50 | 74.92 | 53.61 | 82.44 | 57.37 | 82.11 | 58.25 |
| **Competence** | *AR<50* | 78.26 | 61.69 | 76.5 | 64.81 | 68.42 | 55.77 | 74.92 | 59.81 | 76.05 | 60.78 |
| | *ASCR>5* | 89.79 | 52.69 | 89.62 | 56.49 | 81.59 | 51.04 | 89.59 | 56.44 | 88.61 | 53.65 |
| | *AKPD>7* | 89.72 | 53.67 | 89.4 | 53.26 | 81.87 | 50.74 | 71.77 | 54.43 | 85.14 | 53.50 |
| | *AHS>7* | 88.28 | 54.53 | 88.44 | 56.70 | 81.00 | 52.90 | 86.38 | 51.12 | 87.57 | 54.11 |
| **Familiarity** | *SST>20* | 80.59 | **81.57** | 78.60 | 81.52 | 69.77 | 68.15 | 70.19 | 74.89 | 79.08 | 77.55 |
| | *DST>20* | **81.77** | 72.61 | 75.65 | 69.00 | 70.15 | 57.85 | 72.74 | 66.22 | 64.15 | 68.20 |

Finally, it is also clear that competence is a driver for interaction, but the connection is complex. First, we found that the most frequent pair interactions occur between players on the same squad with relatively high rank (i.e., very competent players). Likewise, pairs that are on average *less* competent have less frequent interaction. Second, we saw a similar story with respect to the absolute difference in rank between the pairs. I.e., players that were of approximately the same rank were increasingly likely to have more interactions together.

ACKNOWLEDGMENT

This research was supported by the U.S. National Science Foundation under Grants No. IIS 1546453. The first author was supported by a scholarship from Jazan University.

REFERENCES

[1] G. A. Benefield, C. Shen, and A. Leavitt. Virtual team networks: How group social capital affects team success in a massively multiplayer online game. In *Proceedings of the 19th ACM Conference on Computer-Supported Cooperative Work & Social Computing*, CSCW '16, pages 679–690, New York, NY, USA, 2016. ACM.

[2] N. V. Chawla, K. W. Bowyer, L. O. Hall, and W. P. Kegelmeyer. Smote: synthetic minority over-sampling technique. *Journal of artificial intelligence research*, 16:321–357, 2002.

[3] A. C. Costa, R. A. Roe, and T. Taillieu. Trust within teams: The relation with performance effectiveness. *European journal of work and organizational psychology*, 10(3):225–244, 2001.

[4] A. C. Cote. i can defend myself. *Games and Culture*, 0(0):1555412015587603, 0.

[5] N. Ducheneaut, N. Yee, E. Nickell, and R. J. Moore. "alone together?": Exploring the social dynamics of massively multiplayer online games. In *Proceedings of the SIGCHI Conference on Human Factors in Computing Systems*, CHI '06, pages 407–416, New York, NY, USA, 2006. ACM.

[6] M. Eftekhar, F. Ronaghi, and A. Saberi. Team formation dynamics: A study using online learning data. In *Proceedings of the 2015 ACM on Conference on Online Social Networks*, COSN '15, pages 257–267, New York, NY, USA, 2015. ACM.

[7] J. A. Espinosa, S. A. Slaughter, R. E. Kraut, and J. D. Herbsleb. Familiarity, complexity, and team performance in geographically distributed software development. *Organization Science*, 18(4):613–630, July 2007.

[8] J. Fox and W. Y. Tang. Womens experiences with general and sexual harassment in online video games: Rumination, organizational responsiveness, withdrawal, and coping strategies. *New Media & Society*, 0(0):1461444816635778, 2016.

[9] C. E. Goodfellow. Online gaming in post-soviet russia: Practices, contexts and discourses. 2015.

[10] P. J. Hinds, K. M. Carley, D. Krackhardt, and D. Wholey. Choosing work group members: Balancing similarity, competence, and familiarity. *Organizational behavior and human decision processes*, 81(2):226–251, 2000.

[11] Y. Huang, W. Ye, N. Bennett, and N. Contractor. Functional or social?: Exploring teams in online games. In *Proceedings of the 2013 Conference on Computer Supported Cooperative Work*, CSCW '13, pages 399–408, New York, NY, USA, 2013. ACM.

[12] M. Hudson, P. Cairns, and A. I. Nordin. *Familiarity in Team-Based Online Games: The Interplay Between Player Familiarity and the Concepts of Social Presence, Team Trust, and Performance*, pages 140–151. Springer International Publishing, Cham, 2015.

[13] K. Kamel, Z. Al Aghbari, and I. Kamel. Realistic team formation using navigation and homophily. In *2014 International Conference on Big Data and Smart Computing (BIGCOMP)*, pages 197–203. IEEE, 2014.

[14] J. Kim, B. C. Keegan, S. Park, and A. Oh. The proficiency-congruency dilemma: Virtual team design and performance in multiplayer online games. In *Proceedings of the 2016 CHI Conference on Human Factors in Computing Systems*, CHI '16, pages 4351–4365, New York, NY, USA, 2016. ACM.

[15] W. A. Mason and A. Clauset. Friends ftw! friendship and competition in halo: Reach. *CoRR*, abs/1203.2268, 2012.

[16] W. Peng and G. Hsieh. The influence of competition, cooperation, and player relationship in a motor performance centered computer game. *Computers in Human Behavior*, 28(6):2100–2106, 2012.

[17] N. Pobiedina, J. Neidhardt, M. d. C. Calatrava Moreno, and H. Werthner. Ranking factors of team success. In *Proceedings of the 22Nd International Conference on World Wide Web*, WWW '13 Companion, pages 1185–1194, New York, NY, USA, 2013. ACM.

[18] M. Ruef, H. E. Aldrich, and N. M. Carter. The structure of founding teams: Homophily, strong ties, and isolation among u.s. entrepreneurs. *American Sociological Review*, 68(2):195–222, 2003.

[19] M. Spante, I. Heldal, A.-S. Axelsson, and R. Schroeder. Strangers and friends in networked immersive environments: Virtual spaces for future living. 2003.

[20] J. C. Waddell and W. Peng. Does it matter with whom you slay? the effects of competition, cooperation and relationship type among video game players. *Computers in Human Behavior*, 38:331–338, 2014.